\documentclass[preprint,showpacs,preprintnumbers,amsmath,amssymb]{revtex4}
\usepackage{graphicx,epsfig,dcolumn,bm,epic,eepic,float}
\usepackage{amsmath}
\usepackage{latexsym}
\usepackage{color}
\usepackage{cancel}
\usepackage{makeidx,shortvrb,latexsym}
\begin{document}
\unitlength 1 cm
\newcommand{\nn}{\nonumber}
\newcommand{\vk}{\vec k}
\newcommand{\vp}{\vec p}
\newcommand{\vq}{\vec q}
\newcommand{\vkp}{\vec {k'}}
\newcommand{\vpp}{\vec {p'}}
\newcommand{\vqp}{\vec {q'}}
\newcommand{\bk}{{\bf k}}
\newcommand{\bp}{{\bf p}}
\newcommand{\bq}{{\bf q}}
\newcommand{\br}{{\bf r}}
\newcommand{\bR}{{\bf R}}
\newcommand{\up}{\uparrow}
\newcommand{\down}{\downarrow}
\newcommand{\fns}{\footnotesize}
\newcommand{\ns}{\normalsize}
\newcommand{\cdag}{c^{\dagger}}

\title{$LHC$  production of forward-center and forward-forward di-jets in the
$k_t$-factorization $unintegrated$ parton distribution frameworks}
\author{
$M. \; Modarres$\footnote{Corresponding author, Email:
mmodares@ut.ac.ir, Tel:+98-21-61118645, Fax:+98-21-88004781}, $M.R.
\; Masouminia$\footnote{Visiting the Institute of Nuclear Physics,
Polish Academy of Science, Krakow, Poland}, $R. \; Aminzadeh\; Nik$}
\affiliation {Department of Physics, University of $Tehran$,
1439955961, $Tehran$, Iran}
\author{$H.\; Hosseinkhani$ }\affiliation{Plasma and Fusion Research School,
Nuclear
Science and Technology Research Institute, 14395-836 Tehran, Iran.}
\begin{abstract}
The present work is devoted to study the high-energy $QCD$ events,
such as the  di-jet productions from proton-proton inelastic
collisions at the $LHC$ in the forward-center and the
forward-forward configurations, using  the $unintegrated$ parton
distribution functions ($UPDF$) in the $k_t$-factorization
framework. The $UPDF$ of $Kimber$ et. al.  ($KMR$) and $Martin$
et.al. ($MRW$) are generated in the leading order ($LO$) and
next-to-leading order ($NLO$), using the $Harland-Lang$ et al.
($MMHT2014$) $PDF$ libraries. While working in the forward-center
and the forward-forward rapidity sectors, one can probe the parton
densities at very low longitudinal momentum fractions ($x$).
Therefore, such a computation can provide a valuable test-field for
these $UPDF$. We find very good agreement with the corresponding
di-jet production data available from $LHC$ experiments. On the
other hand, as we have also stated in our previous works, (i.e. the
protons longitudinal and transverse structure function as well as
hadron-hadron $LHC$ $W/Z$ production),  the present calculations
based on the $KMR$ prescriptions show a better agreement with the
corresponding experimental data. This conclusion is achieved, due to
the particular visualization of the angular ordering constraint ($AOC$),
despite the fact that the $LO-MRW$ and the $NLO-MRW$ formalisms both employ
better theoretical descriptions of the
$Dokshitzer$-$Gribov$-$Lipatov$ -$Altarelli$-$Parisi$ ($DGLAP$)
evolution equation, and hence are expected to produce better
results.  The form of the $AOC$ in the $KMR$ prescription
automatically includes the re-summation of the higher-order $ln({1/x})$ type contributions,
i.e. the $Balitski$-$Fadin$-$Kuraev$-$Lipatov$ ($BFKL$) logarithms,
in the $LO$-$DGLAP$ evolution equation.
\end{abstract}
 \maketitle
\section{Introduction}
Analyzing the raw data, which comes pouring out of the $LHC$,
presents a challenge of considerable proportions, given that the
dynamics of the true players in the hadronic inelastic collisions,
i.e. partons, are shadowed bye the laws of strong interactions.
However, to understand the nature of our universe, it is paramount
to enlighten the behavior of these fundamental substances.
Amazingly, an answer came a few decades ago, in the form of the
$Dokshitzer$-$Gribov$-$Lipatov$-$Altarelli$-$Parisi$ ($DGLAP$)
evolution equations, \cite{DGLAP1,DGLAP2,DGLAP3,DGLAP4}, {
\begin{eqnarray}
    {d \over dlog(Q^2)}g(x,Q^2) &&=  {\alpha_s(Q^2) \over 2\pi}
    \int_x^1 {dz \over z} \left[ P_{gg}^{(LO)}({x \over z})g({z},Q^2)
    + P_{qg}^{(LO)}({x \over z}) \sum_q q({z},Q^2) \right],
    \nonumber \\
    {d \over dlog(Q^2)}q(x,Q^2) &&=  {\alpha_s(Q^2) \over 2\pi}
    \int_x^1 {dz \over z} \left[ P_{qq}^{(LO)}({x \over z})q({z},Q^2)
    + P_{qg}^{(LO)}({x \over z})g({z},Q^2) \right].
    \label{eq1}
\end{eqnarray}
$g(x,Q^2)$ and $q(x,Q^2)$ as the solutions of the $DGLAP$ evolution
equations, are single-scale parton density functions ($PDF$),
corresponding respectively to gluons and quarks.} They depend on the
fraction of the longitudinal momentum of parent hadron ($x$) and an
ultra-violet cutoff ($Q^2$), which denotes the virtuality of the
particle that is being exchanged throughout the inelastic scattering ($IS$).
$P_{ab}^{(LO)}$ are the LO splitting functions (see the section
\ref{secII}). $\alpha_S$ represents the $LO$ running coupling
constant of the strong interaction, conventionally approximated as:
$$
    \alpha_S(Q^2) \simeq {12\pi \over (33-2n_f)log(Q^2/\Lambda_{QCD}^2)},
$$
where $n_f$ is the number of involving flavors in the given strong
interaction and $\Lambda_{QCD}$ is the $QCD$ fundamental low energy
scale. The value of the $\Lambda_{QCD}$ can be effectively extracted
from experiment, around $300$ $MeV$. The terms on the right-hand
side of the equation (\ref{eq1}), correspond to the real emission
and the virtual contributions, respectively.

The main postulation in the $DGLAP$ evolution equation, i.e. the strong ordering hypothesis,
is to neglect the transverse momenta of the partons along the evolution ladder, and to sum over
the $\alpha_S ln(Q^2)$ contributions. One finds out that neglecting the contributions that come
from this transverse dependency may harm the precision of the calculations, particularly in
the high-energy processes and in the small-$x$ region \cite{WMRP,KMR,MRW,KKMS,KIMBER,WattWZ}.
Hence, the need for introducing some transverse momentum dependent ($TMD$) evolution equation
becomes apparent. This gave rise to the $Ciafaloni$-$Catani$-$Fiorani$-$Marchesini$ ($CCFM$) and the
$Balitski$-$Fadin$-$Kuraev$-$Lipatov$ ($BFKL$) evolution equations
\cite{CCFM1,CCFM2,CCFM3,CCFM4,CCFM5,BFKL1,BFKL2,BFKL3,BFKL4,BFKL5}.

One of the main features of the $CCFM$ evolution equation is that it
employs a physical constraint,
to ensure that the gluons emissions are accompanied by constant increase in the angle of
the emission. This feature which is known as the angular ordering constraint ($AOC$),
is related to the color coherent radiations of the gluons
The solutions of the $CCFM$ equation,
$f(x,k_t^2,\mu^2)$ is a double-scaled $TMD$ $PDF$, which in addition
to the $x$ and $Q$, depends on the transverse momentum of the
incoming partons, $k_t$. The idea behind the $CCFM$ evolution
equation (to make the use of the $AOC$ in the evolution ladder) is
valid only in the case of gluon-dominant processes, i.e. in the
small-$x$ sector.
If the proper physical boundaries are inserted, the $CCFM$ equation will
reduce to the conventional $DGLAP$ and $BFKL$ evolutions \cite{CCFM-unfolding}.

Mathematically speaking, solving the $CCFM$ equation is rather
difficult, usually possible with the help of Monte Carlo event
generators, references \cite{MC1,MC2}. On the other hand, the main
feature of the $CCFM$ equation, i.e. the $AOC$, can be used only
for the gluon evolution and therefore, producing convincing
quark contributions in this framework is only a recent development,
see the references \cite{CCFM-Q1,CCFM-Q2,CCFM-Q3}. Given these complexities,
Martin et al, employed the idea of $last-step$ evolution
along the $k_t$-factorization framework, \cite{kt-fact1, kt-fact2,
kt-fact3, kt-fact4, kt-fact6,WMRP}, and developed the
$Kimber$-$Martin$-$Ryskin$ ($KMR$) and the $Martin$-$Ryskin$-$Watt$
($MRW$) approaches \cite{KMR,MRW}. Both of these formalisms are
constructed around the solutions of the $LO$ $DGLAP$ evolution
equations and modified with different visualizations of the angular
ordering constraint. Although the $unintegrated$ parton distribution
functions ($UPDF$) of $MRW$ in the leading order ($LO$) and
next-to-leading order ($NLO$) have been defined to improve the
compatibility of the $KMR$ approach with the theory of the $LO$
$DGLAP$ and extend it to a higher order $QCD$, the recent work
suggests that the $KMR$ framework is more successful (or at least as
successful) in describing experimental data, see for example the
references \cite{Modarres1, Modarres2, Modarres3, Modarres4,
Modarres5, Modarres6, Modarres7, Modarres8, Modarres9}.
Nevertheless, to utter a rigid statement on this matter, further
investigation is required.

One extraordinary test-ground for the $UPDF$ of the
$k_t$-factorization is the probe of the forward-center and
forward-forward rapidity sectors in the hadronic collisions, given
that it involves the dynamics of the small-$x$ region, e.g. $x \sim
10^{-4} - 10^{-5}$, where the gluon density dominates. Since the
decisive difference between the $UPDF$ of $KMR$ and  $MRW$ is in the
different manifestations of the $AOC$, one could argue that working
in such phenomenological setups could potentially exploit this
diversity and unveil the true capacities of the presumed frameworks.
For this propose, we have calculated the process of production of
di-jets in the inelastic proton-proton collisions from the
forward-center and the forward-forward rapidity regions, utilizing the
$UPDF$ of $KMR$ and $MRW$ in the $LO$ and the $NLO$. Comparing these
results with each other, and the results of the similar
calculations in other frameworks, namely the linear and non-linear
$KS$ formalisms, \cite{kutak1, kutak2, kutak3, kutak4, kutak5}, and
with the experimental data from the $CMS$ collaboration \cite{CMS1,
CMS2}, would provide an excellent opportunity to study the strength
and the weaknesses of the $UPDF$ in the $k_t$-factorization framework.

The outlook of this paper is as follows: In the section \ref{secII}
we present a brief introduction to the framework of
$k_t$-factorization and develop the required prescriptions for the
$KMR$ and the $MRW$ $UPDF$, stressing their key differences
regarding the involvement of the $AOC$ in their definitions. The
$UPDF$ will be prepared in their proper $k_t$-factorization schemes
using the $PDF$ of  $Harland-Lang$ et al. ($MMHT2014$) in the $LO$
and the $NLO$, \cite{MMHT}. The section \ref{secIII} contains a
comprehensive description over the utilities and the means for the
calculation of the $k_t$-dependent cross-section of the di-jets
production   in the p-p $IS$ processes. The necessary numerical
analysis will be presented in the section \ref{secIV}, after which a
thorough conclusion will follow in section the \ref{secV}.
\section{The $UPDF$ calculations in the $k_t$-factorization framework}
\label{secII}
During a high energy hadronic
collision, the involving partons, i.e. the partons that appear at
the top of their respective evolution ladders, carry some inherently
induced transverse momentum, as the remnant of the successive (an
potentially infinite) number of evolution steps. When working within
the framework of collinear factorization, such transverse momentum
dependency is conventionally neglected, due to the assumption of the
strong ordering that is embedded in the $LO$ $DGLAP$ evolution
equation,
$$
    k_{t,i-2}^{2} \ll k_{t,i-1}^{2} \ll k_{t,i}^{2} \ll \cdots \ll k_{t,n}^{2} \ll  \mu^2 .
$$
Avoiding such assumption, one can include the contributions coming
from the transverse momentum distributions of the partons, using
either the solutions of the $CCFM$ evolution equation or unify the
$BFKL$ and the $DGLAP$ single-scaled evolution equations to form a
properly tuned $k_t$-dependent framework, \cite{unified,stasto1}.
Utilizing these methods does not always come easy, since these
frameworks are mathematically complex and in the case of $CCFM$, not
enough to include all of the contributing sub-processes.
Alternatively, the single-scaled $PDF$ of the $DGLAP$ evolution
equation can be convoluted with the required $k_t$-dependency during
the last step of the evolution \cite{KIMBER}, postulating that:
$$
    k_{t,i-2}^{2} \ll k_{t,i-1}^{2} \ll k_{t,i}^{2} \ll \cdots \ll k_{t,n}^{2} \sim \mu^2 .
$$
Consequently, one may use the defining identity of the
$k_t$-factorization,
\begin{equation}
    a(x,\mu^2) = \int^{\mu^2} {dk_t^2 \over k_t^2} f_a(x,k_t^2,\mu^2) ,\label{eq6}
\end{equation}
to define the $UPDF$, $f_a(x,k_t^2,\mu^2)$, with $a(x,\mu^2)$ being
the solutions of the $DGLAP$ equation times $x$ (i.e. $xq(x,Q^2)$
and $xg(x,Q^2)$). we should make this comment here  that in the more
precise definition, one should use  the generalized $UPDF$
\cite{kt-fact1, kt-fact2, kt-fact3, kt-fact4,
kt-fact6,WMRP}, i.e. the double-$UPDF$ ($DUPDF$), such that they
take into account both quarks and gluons. Then we should write
(compare with equation (\ref{eq6})):
$$
a(x,\mu^2) = \int^{\mu^2} {dk_t^2 \over k_t^2}\int^1_x
f_a(x,z,k_t^2,\mu^2).
$$
However, in this work we continue our calculations by using the
$UPDF$. Afterwards, one can easily derive the direct expressions for
the $UPDF$ of the $k_t$-factorization, $f_a(x,k_t^2,\mu^2)$.
Furthermore, in order to avoid the soft-gluon singularities, it is
necessary to impose some physical constraint into this definition in
the form of the $AOC$. Naturally, imposing different visualizations
of the $AOC$ will from different formalisms for the $UPDF$.

The first choice is the so called the $KMR$ prescription.
Introducing the virtual (loop) contributions via the Sudakov form
factor,
\begin{equation}
    T_a(k_t^2,\mu^2) = exp \left( - \int_{k_t^2}^{\mu^2} {\alpha_S(k^2) \over 2\pi}
    {dk^{2} \over k^2} \sum_{b=q,g} \int^{1-\Delta}_{0} dz' P_{ab}^{(LO)}(z') \right), \label{eq9}
\end{equation}
and utilizing the $LO$ splitting functions, $P_{ab}^{(LO)}(z=x/x')$,
\begin{eqnarray}
    P_{gg}^{(LO)} (z) &&= \;\; 6 \left( z (1-z) + {1-z \over z} + {z \over 1-z} \right), \nonumber \\
    P_{qq}^{(LO)} (z) &&= \;\; {4 \over 3} \left( {1+z^2 \over 1-z} \right), \nonumber \\
    P_{qg}^{(LO)} (z) &&= \;\;  {1 \over 2} \left( z^2 + (1-z)^2 \right), \nonumber \\
    P_{gq}^{(LO)} (z) &&= \;\; {4 \over 3} {1 + (1-z)^2 \over z},
    \label{eq2}
\end{eqnarray}
as the probability of the emission of a parton $a$ (with the
longitudinal momentum fraction $x$) from a parent parton b (with the
longitudinal momentum fraction $x'$), $Kimber$ et al have defined
the $UPDF$ of $KMR$ as follows:
\begin{equation}
    f_a(x,k_t^2,\mu^2) = T_a(k_t^2,\mu^2) \sum_{b=q,g} \left[ {\alpha_S(k_t^2) \over 2\pi}
    \int^{1-\Delta}_{x} dz P_{ab}^{(LO)}(z) b\left( {x \over z}, k_t^2 \right) \right] . \label{eq8}
\end{equation}
The $LO$ splitting functions parameterize the probability of
evolving from a scale $k_t$ to a higher scale $\mu$ without any
parton emissions. Naturally, the $NLO$ extensions of these functions
would take more complicated forms, see the following equation
(\ref{eq14}) in relation to the $MRW$ prescriptions. The infra-red
cut-off $\Delta=k_t / (\mu + k_t)$ represents a visualization of the
$AOC$, which automatically excludes the $x=x'$ point from the range
of $z$-integration blocking the soft gluon singularities that arise
form the $1/(1-z)$ terms in the splitting functions.

One immediately notes that throughout the above definition, the
$k_t$-dependency gets introduced into the $UPDF$, only at the last
step of the evolution. In order to produce these $UPDF$, the single
scaled $b(x,k_t^2)$ functions can be obtained from the $MMHT2014$
library, \cite{MMHT}, where the calculation of the single-scaled
functions have been carried out using the $IS$ data on the $F_2$
structure function of the proton. Additionally, using the constraint,
$$
    T_a(k_t^2 \geq \mu^2,\mu^2) = 1,
$$
provides the $KMR$ formalism with a smooth behavior over the
small-$x$ region, where the $\alpha_S ln(1/x)$ effects dominate and
the $BFKL$ evolution equation becomes important. The reader should
notice that in the $k_t > \mu$ domain, the unintegrated quark
densities of the $KMR$ approach are non-vanishing, these parton
density functions are considered to be in the $LO$ level.

The second option is the $MRW$ procedure. The $UPDF$ of $KMR$,
despite being proven to have physical value, suffers a
miss-alignment with the theory of the color coherent radiations,
since the $AOC$ is a by-product of the successive gluonic emissions,
therefore, its manifestation (the infra-red cut-off $\Delta$),
should only act on $P_{qq}(z)$ and $P_{gg}(z)$ splitting functions,
i.e. the terms including the on-shell gluon emissions. Correcting
this problem, $Martin$ et al defined the $MRW$ unintegrated
densities in the $LO$ through the following definitions \cite{MRW}
\begin{eqnarray}
    f_q^{LO}(x,k_t^2,\mu^2) &&= T_q(k_t^2,\mu^2) {\alpha_S(k_t^2) \over 2\pi} \int_x^1 dz \left[
    P_{qq}^{(LO)}(z) {x \over z} q \left( {x \over z} , k_t^2 \right) \Theta \left( {\mu \over \mu + k_t}-z \right) \right.
    \nonumber \\
    && \left. + P_{qg}^{(LO)}(z) {x \over z} g \left( {x \over z} , k_t^2 \right) \right], \label{eq10}
\end{eqnarray}
and
\begin{eqnarray}
    f_g^{LO}(x,k_t^2,\mu^2) &&= T_g(k_t^2,\mu^2) {\alpha_S(k_t^2) \over 2\pi} \int_x^1 dz \left[
    P_{gq}^{(LO)}(z)  \sum_q {x \over z} q \left( {x \over z} , k_t^2 \right)
    \right.
    \nonumber \\
    && \left. + P_{gg}^{(LO)}(z) {x \over z} g \left( {x \over z} , k_t^2 \right) \Theta \left( {\mu \over \mu + k_t}-z \right)
    \right], \label{eq11}
\end{eqnarray}
with the modified loop contributions
    \begin{equation}
    T_q(k_t^2,\mu^2) = exp \left( - \int_{k_t^2}^{\mu^2} {\alpha_S(k^2) \over 2\pi} {dk^{2} \over k^2}
    \int^{z_{max}}_{0} dz' P_{qq}^{(LO)}(z') \right), \label{eq12}
    \end{equation}
and
    \begin{equation}
    T_g(k_t^2,\mu^2) = exp \left( - \int_{k_t^2}^{\mu^2} {\alpha_S(k^2) \over 2\pi} {dk^{2} \over k^2}
    \left[ \int^{z_{max}}_{z_{min}} dz' z' P_{gg}^{(LO)}(z')
    + n_f \int^1_0 dz' P_{qg}^{(LO)}(z') \right] \right) , \label{eq13}
    \end{equation}
where $z_{max}=1-z_{min}=\mu/(\mu+k_t)$ \cite{WATT}. To a good
approximation, include the main kinematics of partonic evolution are
included in both of the $UPDF$ of $KMR$ and $MRW$. Interestingly,
the particular choice of the $AOC$ in the $KMR$ formalism, despite
being of the $LO$, includes some higher order contributions, i.e.
from the $ln(1/x)$-dominant sector. On the other hand, in the $MRW$
case, the extension to the higher order must be inserted by the
means of extra constraints.

To include the $NLO$ corrections into the $LO$ $MRW$ framework, one
needs to define the $NLO$ splitting functions as,
\begin{equation}
    \tilde{P}_{ab}^{(LO+NLO)}(z) = \tilde{P}_{ab}^{(LO)}(z) + {\alpha_S \over 2\pi} \tilde{P}_{ab}^{(NLO)}(z),
    \label{eq14}
\end{equation}
with
\begin{equation}
    \tilde{P}_{ab}^{(i)}(z) = P_{ab}^{i}(z) - \Theta (z-(1-\Delta^\prime)) \delta_{ab} F^{i}_{ab} P_{ab}(z),
    \label{eq15}
\end{equation}
with $i=0$ corresponding to the $LO$ and $i=1$ to the $NLO$ levels
(It has been argued that, applying the approximation
$P^{(LO+NLO)}(z) \sim P^{(LO)}(z)$ will simplify the $NLO$
prescription and have a negligible effect on the outcome \cite{MRW},
therefore we do not need to express the exact forms of the $NLO$
splitting functions) . Consequently, the introduction of the $AOC$
into the $NLO$ $MRW$ formalism is through the extended splitting
functions and the $\Theta (z-(1-\Delta^\prime))$ constraint, with
$\Delta^\prime$ being defined as:
$$
    \Delta^\prime = {k\sqrt{1-z} \over k\sqrt{1-z} + \mu}.
$$
Additionally, one have to cut off the tail of the probability into
the $k_t>\mu$ region by inserting a secondary $AOC$ related term
into the body of the real emission sector,
\begin{eqnarray}
    f_a^{NLO}(x,k_t^2,\mu^2) &&= \int_x^1 dz T_a \left( k^2={k_t^2 \over (1-z)}, \mu^2 \right) {\alpha_S(k^2) \over 2\pi}
    \sum_{b=q,g} \tilde{P}_{ab}^{(LO+NLO)}(z)
    \nonumber \\
    &&\times b^{NLO} \left( {x \over z} , k^2 \right) \Theta \left( 1-z-{k_t^2 \over \mu^2} \right).
    \label{eq16}
\end{eqnarray}
The $Sudakov$ form factors in this framework are formulated as:
    \begin{equation}
    T_q(k^2,\mu^2) = exp \left( - \int_{k^2}^{\mu^2} {\alpha_S(q^2) \over 2\pi} {dq^{2} \over q^2}
    \int^1_0 dz' z' \left[ \tilde{P}_{qq}^{(0+1)}(z') + \tilde{P}_{gq}^{(0+1)}(z') \right] \right) ,
    \label{eq17}
    \end{equation}
    \begin{equation}
    T_g(k^2,\mu^2) = exp \left( - \int_{k^2}^{\mu^2} {\alpha_S(q^2) \over 2\pi} {dq^{2} \over q^2}
    \int^1_0 dz' z' \left[ \tilde{P}_{gg}^{(0+1)}(z') + 2n_f\tilde{P}_{qg}^{(0+1)}(z') \right] \right) .
    \label{eq18}
    \end{equation}
The reader can find a comprehensive description of the $NLO$
splitting functions in the references \cite{MRW,PNLO}.

\begin{figure}[ht]
\includegraphics[scale=0.3]{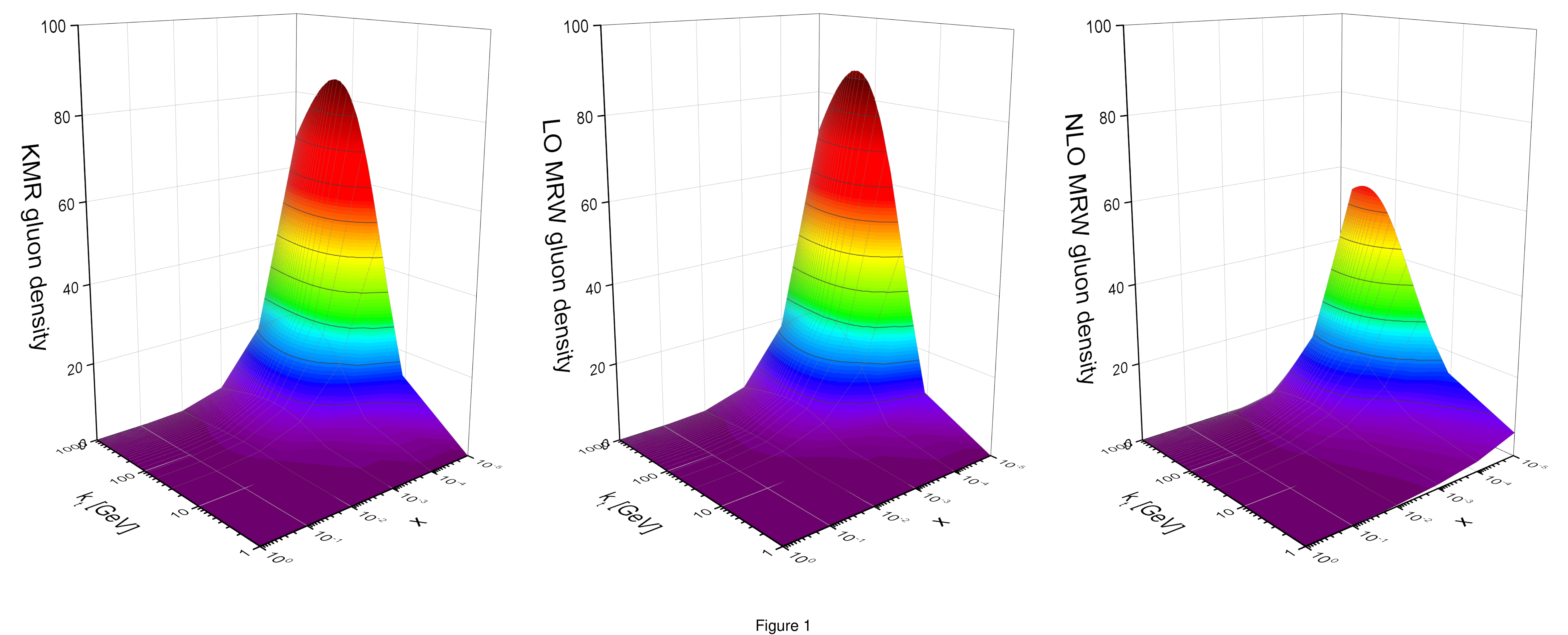}
\caption{The gluonic $UPDF$ of the $k_t$-factorization versus
 the fractional longitudinal momentum of the parent hadron
($x$) and the transverse momentum of the parton, appearing on the
top of the evolution ladder ($k_t$) at $\mu = 100 \; GeV$. The
difference in the behavior of the $UPDF$ in different frameworks is
a direct consequence of employing different manifestations of the
$AOC$ in their respective definitions. To plot these diagrams we
have used the $PDF$ libraries of $MMHT2014$ in the $LO$ and the
$NLO$ as the input for the equations (\ref{eq8}), (\ref{eq11}) and
(\ref{eq16}).} \label{fig1}
\end{figure}

In the  figure \ref{fig1}, the $UPDF$ of the $k_t$-factorization are
plotted against the fractional longitudinal momentum of the parent
hadron ($x$) and the transverse momentum of the parton, appearing on
the top of the evolution ladder ($k_t$). The obvious difference in
the behavior of the $UPDF$ in different frameworks is a direct
consequence of employing different manifestations of the $AOC$ in
their respective definitions.
\section{The Di-jet production in the p-p collisions at the $LHC$}
\label{secIII} Generally speaking, the main contributions into the
hadronic cross-section of the di-jet productions at the $LHC$, i.e.,
$$ P_1 + P_2 \to J_1 + J_2 + X, $$
are the $LO$ partonic sub-processes:
\begin{eqnarray}
    g(\textbf{k}_1) + g^{*}(\textbf{k}_2) \to g(\textbf{p}_1) + g(\textbf{p}_2), \nonumber \\
    g(\textbf{k}_1) + g^{*}(\textbf{k}_2) \to q(\textbf{p}_1) + \bar{q}(\textbf{p}_2), \nonumber \\
    q(\textbf{k}_1) + g^{*}(\textbf{k}_2) \to q(\textbf{p}_1) + g(\textbf{p}_2).
    \label{eq19}
\end{eqnarray}
Since we are considering the forward sector for the partons that are
produced in the $k_t$-factorization, the stared partons in the
equation (\ref{eq19}), one can safely neglect the $qq$ and
$q\bar{q}$ sub-processes.
 In the collinear factorization framework, the cross-section of
a hadronic $IS$ can be written as a sum over all of the involving
partonic cross-sections, times the probability of appearing the
particular partonic configuration at top of the evolution ladder of
the individual hadrons, i.e.,
\begin{eqnarray}
    \sigma_{Hadron-Hadron} &&= \sum_{a_1,a_2=q,g} \int_0^1 {dx_1 \over x_1} \int_0^1 {dx_2 \over x_2}
    \; a_1(x_1,\mu_1^2)\; a_2(x_2,\mu_2^2) \; \nonumber \\[0.4cm]
    && \times \hat{\sigma}_{a_1- a_2}(x_1,k^2_{1,t}=0,\mu_1^2;x_2,k^2_{2,t}=0,\mu_2^2),\nonumber\\
    \label{eq20}
\end{eqnarray}
where $\hat{\sigma}_{a_1- a_2}$ denotes the cross-section of the
incoming partons $a_1$ and $a_2$, respectively with the longitudinal
momentum fractions $x_1$ and $x_2$, the hard scales $\mu_1$ and
$\mu_2$ and neglected transverse momenta. $\hat{\sigma}_{a_1- a_2}$
may be defined as follows:
\begin{equation}
    d\hat{\sigma}_{a_1 a_2} = d\phi_{a_1 a_2} {|{\mathcal{M}_{a_1 a_2}}|^2 \over F_{a_1 a_2}},
    \label{eq21}
\end{equation}
with the multi-particle phase space $d\phi_{a_1 a_2}$,
    \begin{equation}
    d\phi_{a_1 a_2} \equiv \prod_i {d^3 p_i \over 2E_i} \delta^{(4)} \left( \sum p_{in} - \sum p_{out} \right) ,
    \label{eq22}
    \end{equation}
and the flux factor $F_{a_1 a_2}$,
    \begin{equation}
    F_{a_1 a_2} \equiv x_1 x_2 s.
    \label{eq23}
    \end{equation}
$s$ is the center of mass energy squared,
    $$ s=(P_1 + P_2)^2=2P_1.P_2, $$
with $P_1$ and $P_2$ being the 4-momenta of the incoming hadrons,
where we have neglected the mass of the proton, while working in the
infinite momentum frame. ${\mathcal{M}_{a_1 a_2}}$ in the equation
(\ref{eq21}) are the matrix elements of the partonic sub-processes,
the equations (\ref{eq19}). To calculate these quantities, one must
first understand the exact kinematics that rule over the
corresponding partonic sub-processes.

To include the contributions coming from the transverse momentum
dependency of the probability functions, one can use the definition
of the $UPDF$ in the framework of $k_t$-factorization, the equation
(\ref{eq6}) and rewrite the equation (\ref{eq20}) as follows:
\begin{eqnarray}
    \sigma_{Hadron-Hadron} &&= \sum_{a_1,a_2=q,g} \int_0^1 {dx_1 \over x_1} \int_0^1 {dx_2 \over x_2}
    \int_{0}^{\infty} {dk^2_{1,t} \over k^2_{1,t}} \int_{0}^{\infty} {dk^2_{2,t} \over k^2_{2,t}}
    \; f_{a_1}(x_1,k^2_{1,t},\mu_1^2) \; f_{a_2}(x_2,k^2_{2,t},\mu_2^2)
    \nonumber \\[0.4cm] &&
    \times \hat{\sigma}_{a_1 a_2}(x_1,k^2_{1,t},\mu_1^2;x_2,k^2_{2,t},\mu_2^2).
    \label{eq24}
\end{eqnarray}
Now, it is convenient to characterize $d\phi_{a_1 a_2}$ in term of
the transverse momenta of the product particles, $p_{i,t}$, their
rapidities, $y_i$, and the azimuthal angles of the emissions,
$\varphi_i$,
\begin{equation}
    {d^3 p_i \over 2E_i} = {\pi \over 2} dp_{i,t}^2 dy_i {d\varphi_i \over 2\pi}.
    \label{eq25}
\end{equation}
Working in the proton-proton center of mass frame, one may use below kinematics,
\begin{eqnarray}
    P_1 &&= {\sqrt{s} \over 2} (1,0,0,1), \;\;\; P_2 = {\sqrt{s} \over 2} (1,0,0,-1),
    \nonumber \\
    \textbf{k}_i &&= x_i \textbf{P}_i + \textbf{k}_{i,\perp}, \;\;\; k_{i,\perp}^2 = -k_{i,t}^2,  \;\;\; i=1,2 \; ,
    \label{eq26}
\end{eqnarray}
where the $k_i$ are the 4-momenta of the partons that enter the
semi-hard process. Then, for each partonic sub-process, the conservation of
the transverse momentum reads as,
\begin{equation}
    \textbf{k}_{1,\perp} + \textbf{k}_{2,\perp} = \textbf{p}_{1,\perp} + \textbf{p}_{2,\perp}.
    \label{eq27}
\end{equation}
Afterwards, one can simply define,
\begin{eqnarray}
    x_1 &&= {1 \over \sqrt{s}} \left( p_{1,t} e^{+y_1} + p_{2,t} e^{+y_2} \right),
    \nonumber \\
    x_2 &&= {1 \over \sqrt{s}}  \left( p_{1,t} e^{-y_1} + p_{2,t} e^{-y_2} \right).
    \label{eq28}
\end{eqnarray}

\begin{figure}[ht]
\includegraphics[scale=0.35]{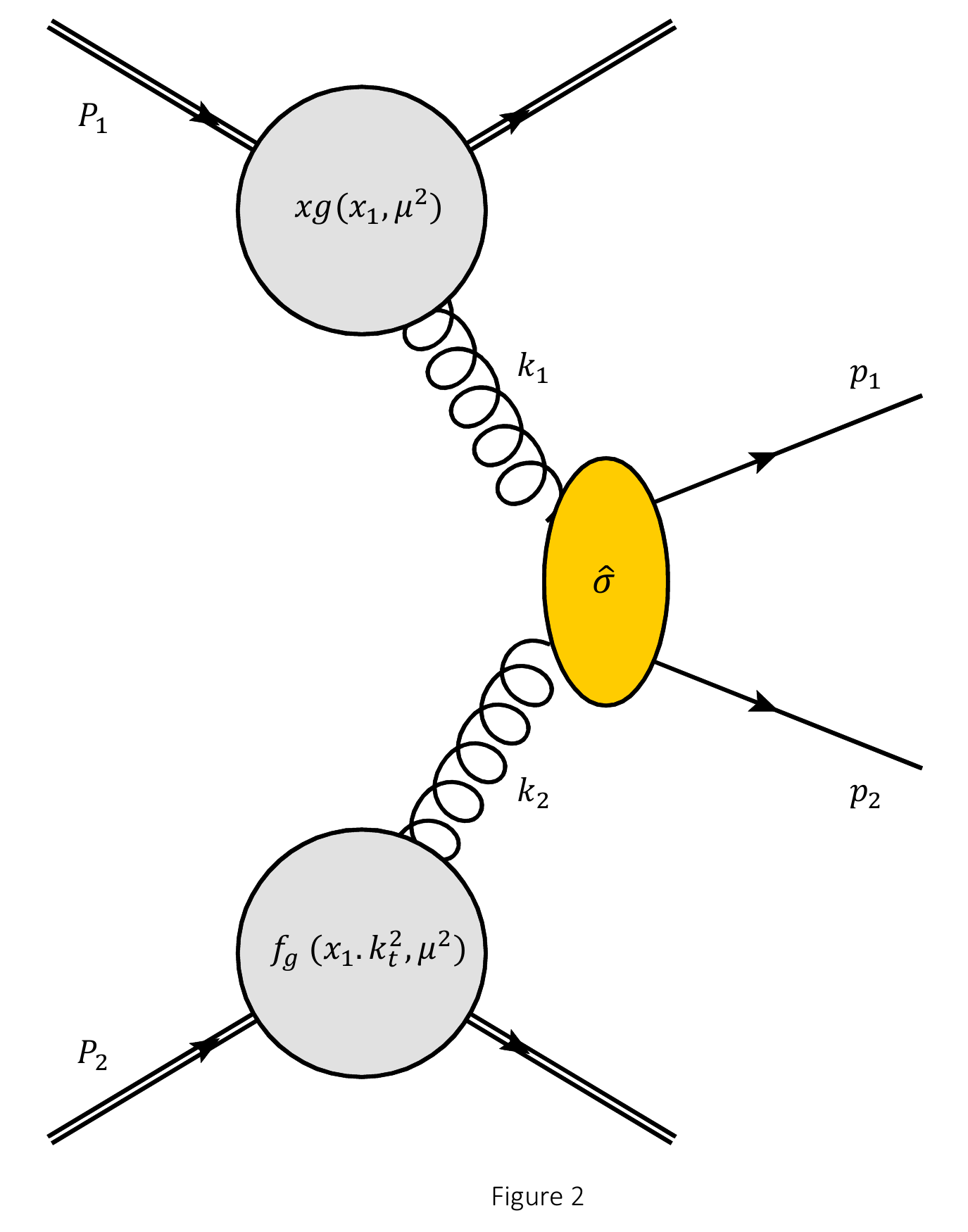}
\caption{The deep inelastic scattering of two protons in the
forward-center configuration. The diagram shows the $g^{*}+g \to
q+\bar{q}$ sub-process, assuming that one of the quarks is being
produced in the forward sector (bounded by $3.2 < |\eta_{f}| < 4.7$)
and the other in the center sector (bounded by $|\eta_{c}| < 2.8$).
The parton density related to the first proton is being described
with the $integrated$ $PDF$ while the second parton is prepared
using the $UPDF$ in one of our presumed frameworks.} \label{fig2}
\end{figure}

The figure \ref{fig2} illustrates the schematics for a proton-proton
deep inelastic collision in the forward-center (or the
forward-forward) rapidity sector in a particular partonic
sub-process, i.e. $g^{*}+g \to q+\bar{q}$. Working within the
boundaries of the forward-center or the forward-forward rapidity
sector, without damaging the main assumptions, one can assume that
$x_1 \sim 1$ and $x_2 \ll 1$. In the direct consequent of a such
approximation, we can safely neglect the transverse momentum
dependency of the first parton entering the hard process (shift it
to the collinear domain), and rewrite the equation (\ref{eq24}) as,
\begin{eqnarray}
    \sigma_{Hadron-Hadron} &&= \sum_{a_1,a_2=q,g} \int_0^1 {dx_1 \over x_1} \int_0^1 {dx_2 \over x_2}
    \int_{0}^{\infty} {dk^2_{t} \over k^2_{t}}
    a_1(x_1,\mu_1^2) \; f_{a_2}(x_2,k^2_{t},\mu_2^2)
    \nonumber \\[0.4cm] &&
    \times \hat{\sigma}_{a_1 a_2}(x_1,\mu_1^2;x_2,k^2_{t},\mu_2^2),
    \label{eq29}
\end{eqnarray}
with the $k_t$ being defined as,
\begin{equation}
    k_t = \left[ p_{1,t}^2 + p_{2,t}^2 + 2 p_{1,t} p_{2,t} cos(\Delta \varphi) \right]^{1/2},
    \label{eq30}
\end{equation}
and $\Delta \varphi = \varphi_1 - \varphi_2$.

After determining the kinematics of the involving processes, it is
possible to calculate their matrix elements, i.e. ${\mathcal{M}_{a_1
a_2}}$. To this end, one have to sum over the $dk_{i,t}^2/k_{i,t}^2$
terms only from the ladder-type diagrams, and somehow systematically
dispose  the interference (the non-ladder) diagrams, e.g. by using a
physical gauge for the gluons,
    \begin{equation}
    d_{\mu \nu} (k) = -g_{\mu \nu} + { k_{\mu} n_{\nu} + n_{\mu} k_{\nu} \over k.n } .
    \label{eq31}
    \end{equation}
Note that $n = x_1 P_1 + x_2 P_2$ is the gauge-fixing vector. One
might expect that neglecting the contributions coming from the
non-ladder diagrams, i.e. the diagrams where the production of the
jets is a by-product of the hadronic collision (see the reference
\cite{Modarres9,Deak1}), would have a numerical effect on the
results. Hence, using the equation (\ref{eq31}) as our choice for the axial
gauge for the gluons, we can safely subtract the "unfactorizable" contributions
coming from the non-ladder type diagrams. Thus, using the regular Feynman rules, inserting
the "non-sense" polarization for the incoming gluons
\begin{equation}
    \sum \epsilon^{\mu} (\textbf{k}_i) \epsilon^{* \nu} (\textbf{k}_i) =
    { k_{i,t}^{\mu} k_{i,t}^{\nu} \over \textbf{k}_{i,t}^2},
    \label{eq32}
\end{equation}
and imposing the "eikonal" approximation to justify the use of an
on-shell prescription for the off-shell particles (via neglecting
the exchanged momenta in the quark-gluon vertices and preserving the
spin of the gluons, see the references \cite{Modarres9,Deak1,LIP1}),
\begin{equation}
    -i \bar{u}(p_i) \gamma^{\mu} u(p_i) \to {-2i \over k_{i,t}^2} P_i^{\mu},
    \label{eq33}
\end{equation}
one can manage to extract the matrix element, corresponding to the
processes of the equation (\ref{eq19}), see the appendix A.

Now, using the above equations, one can derive the master equation
for the total cross-section of the production of di-jets in the
framework of $k_t$-factorization,
\begin{eqnarray}
    \sigma_{p-p}(P_1+P_2 \to J_1+J_2)  &&= \sum_{a,c,d=q,g} {1 \over 1 + \delta_{cd}} \int
    {p_{1,t} p_{2,t} \over 8\pi^2 (x_1 x_2 s)^2}
    dy_1 dy_2 {dp_{1,t} dp_{2,t} \over  k^2_{t}} d\Delta \varphi \; a(x_1,\mu^2)
    \nonumber \\[0.4cm] &&
    \times \; f_{g}(x_2,k^2_{t},\mu^2)  \;
    |\mathcal{M}_{a+g \to c+d}(x_1,\mu^2;x_2,k^2_{t})|^2.
    \label{eq34}
\end{eqnarray}
The term $1 /(1 + \delta_{cd})$ restrains the over-counting indices.
Note that, the existence of the term $k^{-2}_{t}$ in the equation
(\ref{eq34}) is the remnant of the re-summation factor,
$dk_{t}^2/k_{t}^2$, from the equation (\ref{eq6}) and since we are
interested to look for the transverse momentum dependent jets with
$p_{i,t} > 20 \; GeV$, the presence of such denominator would not
cause any complication in the master equation. Additionally, we have
to decide how to validate our $UPDF$ in the non-perturbative region.
i.e. where $k_t<\mu_0$ with $\mu_0=1\;GeV$. A natural option would
be to fulfill the requirement that:
    $$
    \lim_{k_{t}^2 \rightarrow 0} f_g(x,k_{t}^2,\mu^2) \sim k_{t}^2,
    $$
and therefore, one can safely choose the following approximation for
the non-perturbative region:
    \begin{equation}
    f_g(x,k_{t}^2<\mu_0^2,\mu^2) = {k_{t}^2 \over \mu_0^2} g(x,\mu_0^2) T_g(\mu_0^2,\mu^2).
    \label{eq35}
    \end{equation}

In the next section, we will introduce some of the numerical methods
that have been used for the calculation of the cross-section of the
production of di-jets, using the $UPDF$ of $KMR$ and $MRW$.
\section{The numerical analysis}
\label{secIV} We perform the 5-fold integration of the master
equation (\ref{eq34}), using the $\mathtt{VEGAS}$ algorithm in
Monte-Carlo integration. To do this, we have selected the hard-scale
of the $UPDF$ as the share of each of the parent hadrons from the
total energy of the center-of-mass frame:
\begin{equation}
    \mu = {1 \over 2}E_{CM}.
    \label{eq36}
\end{equation}
Variating this normalization value around a factor of 2, will
provide each framework with a decent uncertainty bound. One would
also set the upper boundaries on the transverse momentum
integrations to $ p_{i,max} = 4 \mu$, noting that increasing this
upper value does not have any effect on the outcome.

The forward rapidity sectors is conventionally defined as,
\begin{equation}
    3.2 < |\eta_{f}| < 4.8,
    \label{eq37}
\end{equation}
where $\eta$ denotes the pseudorapidity of a produced particle,
$$ \eta = - ln \left[ tan \left( {\theta \over 2} \right) \right], $$
with $\theta$ being the angle between the propagation axis and the
momentum of the particle. Alternatively, to work in the central
rapidity sector, one have to choose,
\begin{equation}
    |\eta_{c}| < 2.8.
    \label{eq38}
\end{equation}
Therefore, while working in the infinite momentum frame i.e. where
$\eta \simeq y$, to perform our calculations in the forward-center
region, we set:
$$ y_1 =  \eta_{c}, \;\;\; y_2 =  \eta_{f}. $$
Trivially, the choice
$$ y_1 =  \eta_{f}, \;\;\; y_2 =  \eta_{f}, $$
marks the forward-forward region. Such framework should be ideal to
describe the inclusive $CMS$ data regarding the forward-center
di-jet measurements for $p_{i,t}>35\;GeV$. After confirming that,
one can go further, producing predictions in the framework of
forward-forward di-jet production for the $LHC$.

Moreover, as a consequence of employing the inclusive scenario (i.e.
$p_{i,t}>35\;GeV$ and limiting the rapidity integrations to the
forward or central regions), one must assure that the produced jets
must lie within this specific region. Thus, in order to cut-off the
collinear and the soft singularities, it is conventional to use the
anti-$k_t$ algorithm \cite{Cacciari}, with radius $R = 1/2$,
bounding the jets to this particular initial setup, through
inserting a constraint on the $y-\varphi$ plane:
\begin{equation}
    R  > \left[ (\Delta \varphi)^2 + (y_2-y_1)^2 \right]^{1/2}.
    \label{eq39}
\end{equation}
Introducing the anti-$k_t$ jet constraint ensures the production of
2 separated jets and rejects any single-jet scenarios.

\section{Results, Discussions and Conclusions}
\label{secV} Having in mind the theory and the notions of the
previous sections, we are able to calculate the production rates
belonging to the di-jets in the forward-center and the
forward-forward rapidity sectors, from the perspective of the
$k_t$-factorization framework, utilizing the $UPDF$ of $KMR$ and
$MRW$. The $PDF$ of $Harland-Lang$ et al. \cite{MMHT}, $MMHT2014$,
in the $LO$ and $NLO$ levels, are used as the input functions for the
unintegrated gluon densities, i.e., the equations (\ref{eq8}),
(\ref{eq11}) and (\ref{eq16}). Additionally, they are fit to be used
as the solutions of the $DGLAP$, the $PDF$ of the collinear
factorization, directly in the master equation (\ref{eq34}). We tend
to perform the above calculations in any of our presumed frameworks,
the $KMR$, the $LO$ $MRW$ and the $NLO$ $MRW$, then compare the
results to each other, to the similar calculations in other
frameworks and to the existing experimental data, in the case of the
forward-center.

\begin{figure}[ht]
\includegraphics[scale=0.35]{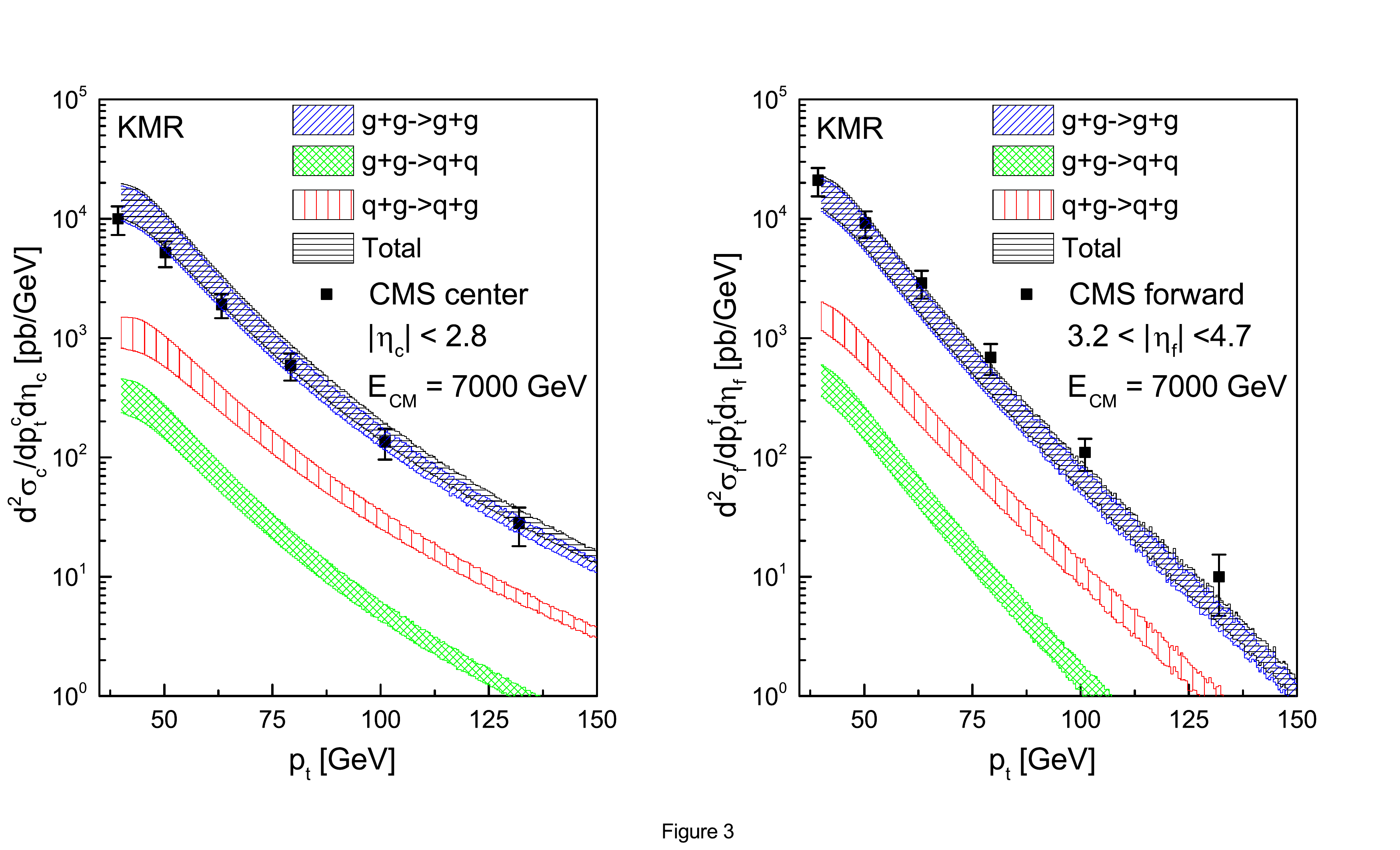}
\caption{The differential cross-section for the production of
di-jets in the forward-center rapidity sector, calculated in the
$KMR$ framework for $E_{CM} = 7\;TeV$. The contributions from each
of the involving sub-processes form the equation (\ref{eq19}) have
been plotted separately. The black-oblique patterned histograms
illustrate the sum of the partonic contributions. To determine the
uncertainty of the calculations, we have manipulated the hard scale
of the $UPDF$, $\mu = E_{CM}/2$, by a factor of 2. The data point
are from the measurements of the $CMS$ collaboration, the reference
\cite{CMS1}.} \label{fig3}
\end{figure}

\begin{figure}[ht]
\includegraphics[scale=0.35]{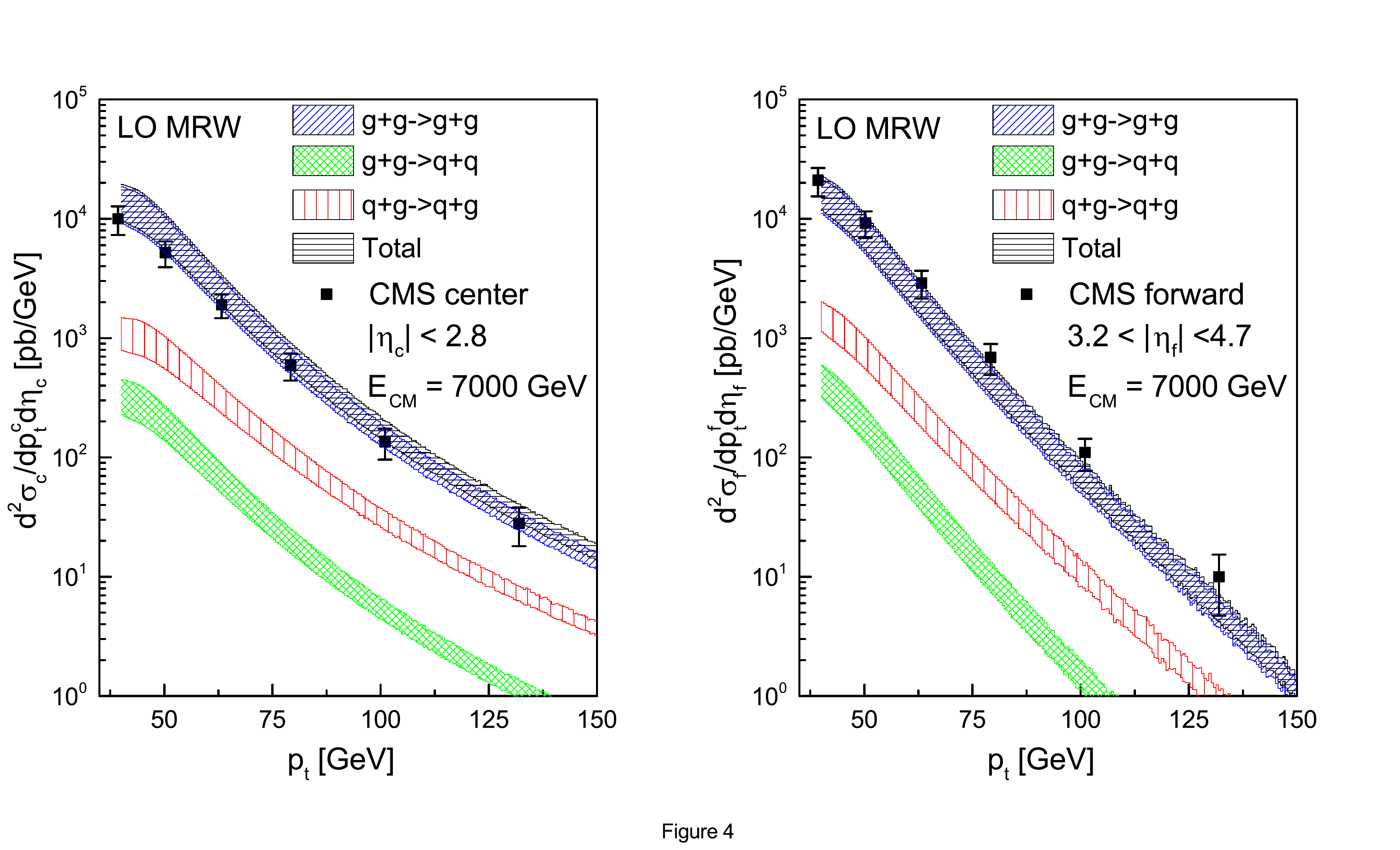}
\caption{The differential cross-section for the production of
di-jets in the forward-center rapidity sector, calculated in the
$LO$ $MRW$ framework. The notion of the diagrams are as in the
figure \ref{fig3}.} \label{fig4}
\end{figure}

\begin{figure}[ht]
\includegraphics[scale=0.35]{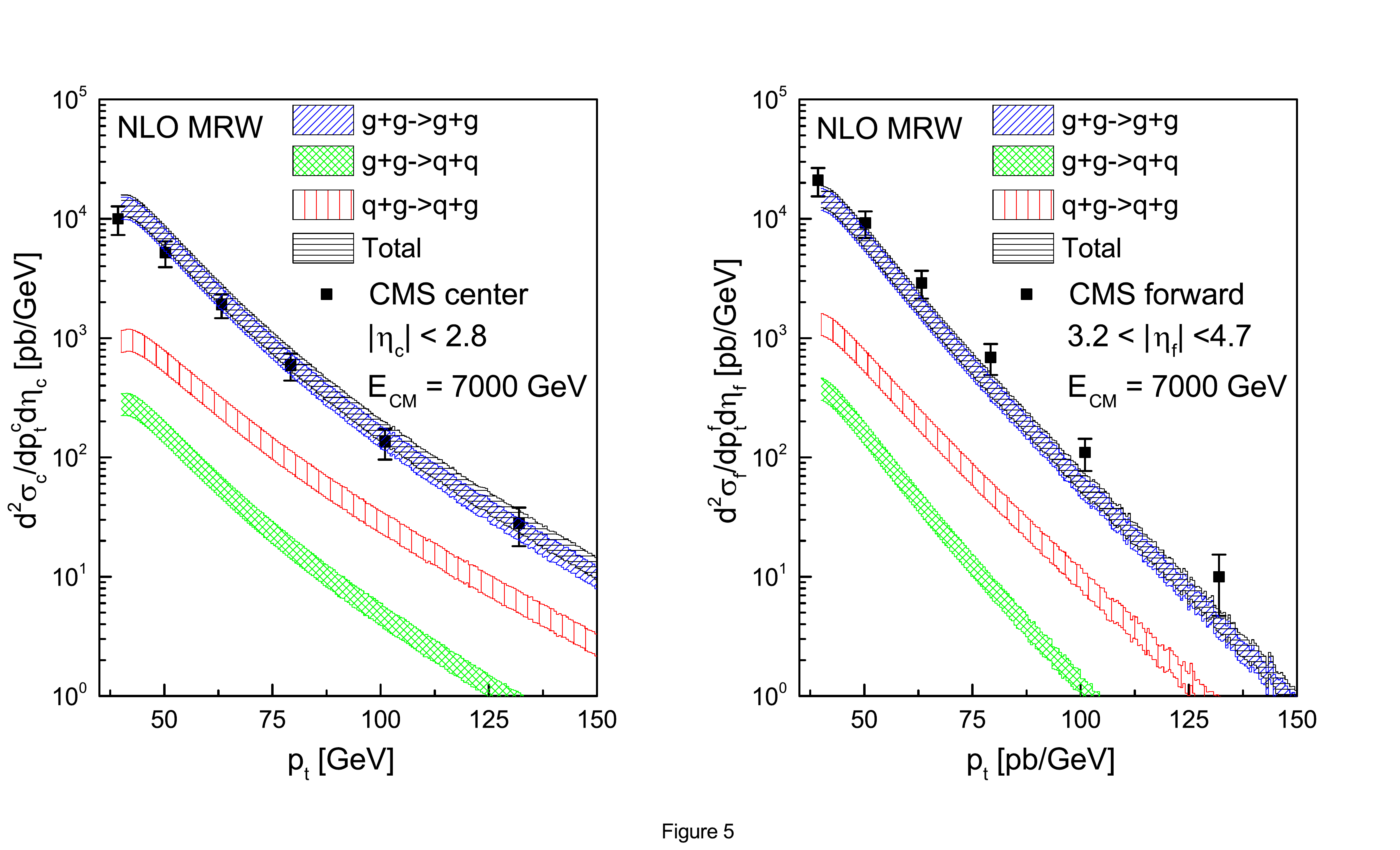}
\caption{The differential cross-section for the production of
di-jets in the forward-center rapidity sector, calculated in the
$NLO$ $MRW$ framework. The notion of the diagrams are as in the
figure \ref{fig3}.} \label{fig5}
\end{figure}

\begin{figure}[ht]
\includegraphics[scale=0.35]{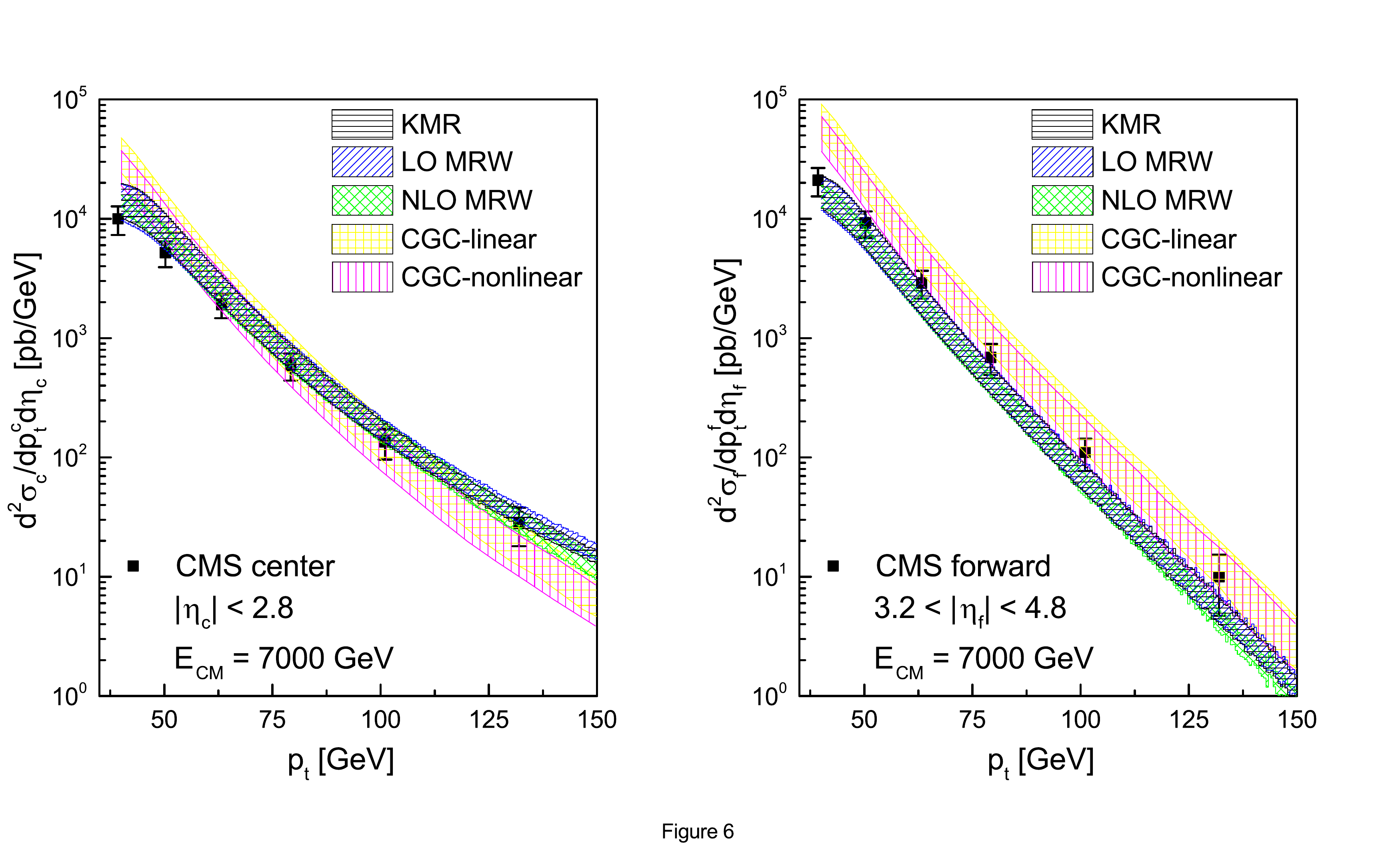}
\caption{The comparison between the differential cross-sections of
the production of di-jets from the forward-center rapidity sector,
in the different frameworks of the $k_t$-factorization. The results
have been prepared as the numerical solutions the equation
(\ref{eq34}), using the $UPDF$ of $KMR$ and $MRW$ in the $LO$ and
$NLO$ with $E_{CM} = 7\;TeV$. The data points are from the $CMS$
report \cite{CMS1}. The yellow-checkered and the purple-vertically
stripped patters represent the calculations in the linear and
non-linear $KS$ frameworks, respectively, see the reference
\cite{kutak1}.} \label{fig6}
\end{figure}

\begin{figure}[ht]
\includegraphics[scale=0.35]{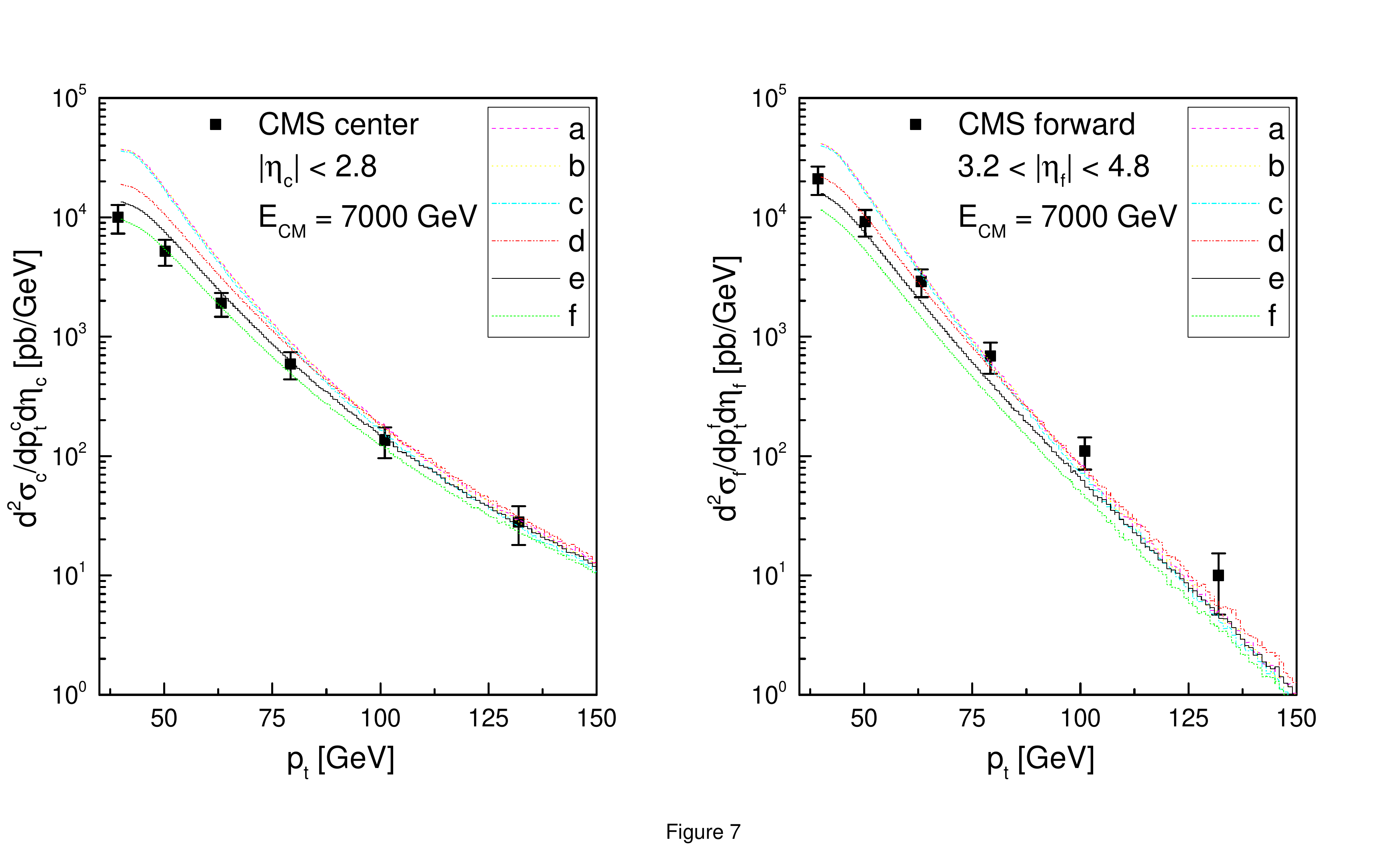}
\caption{The differential cross-section for the production of
di-jets in the forward-center rapidity sector, for different choices
of the hard scale and from the dominant $g^{*}+g \to g + g$
sub-process. The calculations have been carried on in the $KMR$
framework for $E_{CM} = 7\;TeV$. The histograms $a$ through $f$ have
been calculated using the conditions from the equation (\ref{eq40}).
We have chosen the condition $e$ (the black-continues histograms),
i.e. the equation (\ref{eq36}), as the primary prescription
throughout this work.} \label{fig7}
\end{figure}

So, the figures \ref{fig3}, \ref{fig4} and \ref{fig5} present the
reader with the differential cross-section for the production of
well-separated forward-central di-jets ($d^2\sigma/dp_t d\eta$),
plotted against the transverse momentum of the corresponding jets
($p_t$) in the $KMR$, the $LO$ $MRW$ and the $NLO$ $MRW$ schemes
respectively. The uncertainty bounds are calculated, variating the
hard scale of the $UPDF$ with a factor of 2, since this is the only
arbitrary physical parameter in the framework of
$k_t$-factorization. The blue-hatched pattern, the green-checkered
and the red-vertically stripped patterns illustrate the individual
contributions of the partonic sub-processes from the equation
(\ref{eq19}), corresponding to the $g^{*}+g \to g+g$, $g^{*}+g \to
q+\bar{q}$ and $g^{*}+q \to g+q$ processes respectively. The
black-horizontally stripped pattern represents the sum of the
sub-contributions. The calculations have been compared against the
experimental data of the $CMS$ collaboration, the reference
\cite{CMS1}. One immediately notices that the share of the $g^{*}+g
\to g+g$ sub-process dominates, relative to the negligible shares of
the remaining two sub-processes. Although all of these frameworks
are relatively successful in describing the experimental data, see
the figure \ref{fig6}, it is interesting to find that the $UPDF$ of
$KMR$ do as well as (if not better than) the $UPDF$ of $MRW$ in
predicting the experimental results. The closeness of the behavior
of different frameworks is a consequence of our choice for the hard
scale of the $UPDF$, the equation (\ref{eq36}). In order to enlighten
this point, the figure \ref{fig7} illustrates the result of making
different choices in such calculations, using the $UPDF$ of the
$KMR$. To demonstrate the effect of changing the hard scale of the
$UPDF$ in the outcome, the histograms are calculated utilizing the
following hard scale prescriptions
\begin{eqnarray}
    &&a)  \; \; \; \mu = {1 \over 2}\left( p_{1,t} + p_{2,t} \right), \nonumber \\
    &&b)  \; \; \; \mu = {1 \over 2}\left( p_{1,t}^2 + p_{2,t}^2 \right)^{1/2}, \nonumber \\
    &&c)  \; \; \; \mu = Max(p_{1,t},p_{2,t}), \nonumber \\
    &&d)  \; \; \; \mu = {1 \over 4}E_{CM}, \nonumber \\
    &&e)  \; \; \; \mu = {1 \over 2}E_{CM}, \nonumber \\
    &&f)  \; \; \; \mu = E_{CM},
    \label{eq40}
\end{eqnarray}
where $ Max(p_{1,t},p_{2,t})$ returns the higher value between the
transverse momenta of the produced jets. To save computation time,
we only considered the contributions coming form the dominant
$g^{*}+g \to g + g$ sub-processes. The choice $a$, which have been
used in the similar calculations (e.g., the references
\cite{kutak1,kutak2,kutak3,kutak4,kutak5} in the high energy
factorization, from the point of view of the $UPDF$ of the color
gloss condensation, ($CGC$)) proves to be in contrast with the
particular manifestation of the $AOC$, specially in the case of
$NLO$ $MRW$ $UPDF$. This is in addition to the considerable
off-shoot of the results in the smaller values of the transverse
momenta belonging to the produced jets. In the figure \ref{fig6},
the yellow-checkered and the purple-vertically stripped patters
represent the calculations in the linear and the non-linear $KS$
frameworks, respectively. The above separation between the
predictions of the $KS$ framework and the experimental data is
apparent. To avoid such complications, we have chosen the condition
$e$, in the equation (\ref{eq40}), as the primary prescription for the hard
scale of our $UPDF$ throughout this work, see the section
\ref{secIV}.

Having a closer look into the figure \ref{fig6}, one notices that
such off-shooting results also appear in our settings for the
production of di-jets. This is perhaps because of the
over-simplified dynamics that have been used to derive these
measurements. An increase in the precision may be realized via
including higher order diagrams and introducing the final state
parton showers in this frameworks \cite{kutak6}. Beside this point,
note that our results show an acceptable agreement with the
experimental data of the $CMS$ collaboration, reference \cite{CMS1}.
Another interesting observation is that in the large $k_t$, where
the higher order corrections become important, the calculations in
the $KMR$ approach start to separate from the $LO \; MRW$ and behave
similar to the $NLO \; MRW$. The reason is that the inclusion of the
$non-diagonal$ splitting functions into the domain of the $AOC$
introduces some corrections from the $NLO$ region (in the form of
$ln(1/x)$ re-summations) into the $KMR$ formalism.

\begin{figure}[ht]
\includegraphics[scale=0.35]{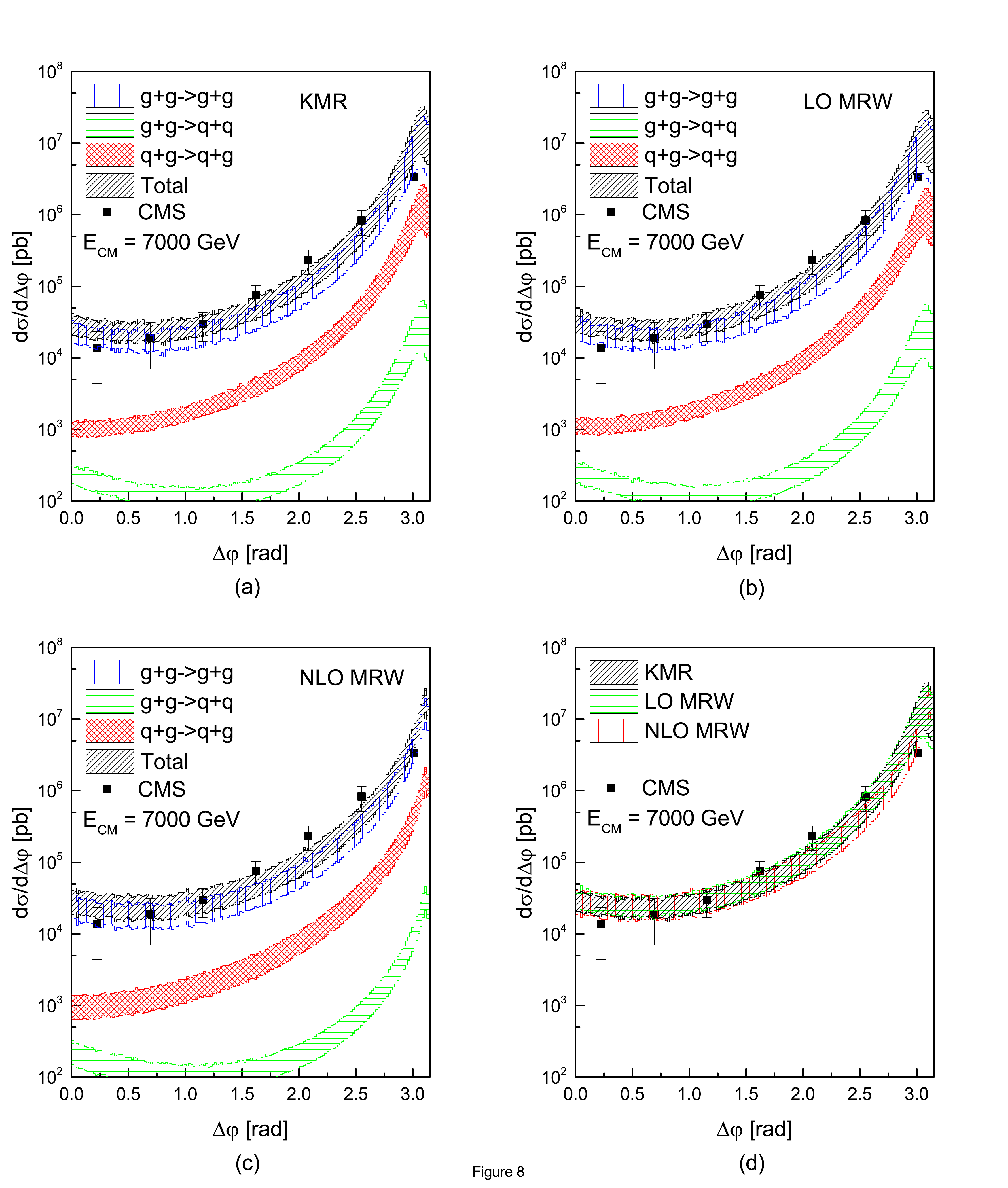}
\caption{The differential cross-section for the production of
di-jets versus the angle of the out-coming jets, $\Delta \varphi$.
The calculations are in the forward-center rapidity sector for
$E_{CM} = 7\;TeV$. The panels (a), (b) and (c) illustrate the
calculations, utilizing the $UPDF$ of $KMR$, $LO$ $MRW$ and $NLO$
$MRW$, respectively. The contributions from each of the involving
sub-processes are shown separately. The panel (d) presents the
comparison of these measurements against each other as well as the
experimental data of the $CMS$ collaboration, the reference
\cite{CMS2}. The uncertainty of the calculations are provided
through manipulating the hard scale of the $UPDF$ by a factor of 2.}
\label{fig8}
\end{figure}

A recent report from the $CMS$ collaboration, the reference
\cite{CMS2}, concerns the angular distribution of the produced jets
in the forward-center rapidity sector from a deep inelastic event at
the $LHC$. Making use of this new information, we have calculated
the differential cross-section of the forward-central di-jet
production ($d\sigma/d \Delta \varphi$), plotted in the figure
\ref{fig8} against the angular difference of the produced partons
(or equivalently the angular difference of the produced jets,
$\Delta \varphi$). The panels   (a), (b) and (c) in this figure
illustrate the details of the calculations in each framework,
consisting of the individual contributions of the sub-processes and
the corresponding uncertainty bounds. The panel (d) presents the
reader with the comparison of the total amounts in the presumed
formalisms to each other and to the data from the reference
\cite{CMS2}. Again, the results in the $KMR$ approach seems to be
equally good (or better than) those from the $MRW$ in the $LO$ or
the $NLO$.

\begin{figure}[ht]
\includegraphics[scale=0.35]{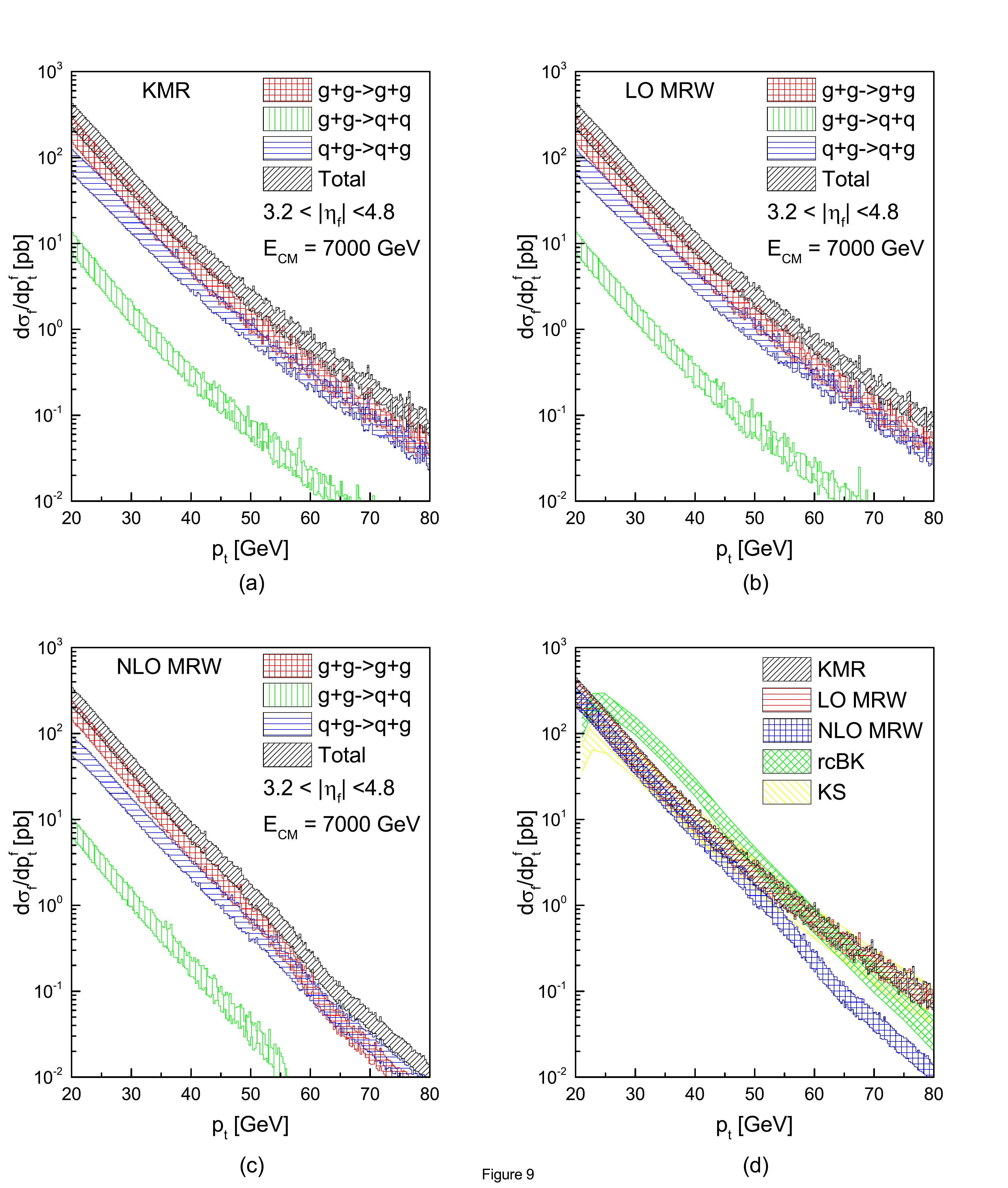}
\caption{The calculated predictions for the production of
forward-forward di-jets in the framework of $k_t$-factorization with
the central-mass energy of $7\;TeV$. The differential cross-section
for the production of di-jets are plotted against the transverse
momenta of the produced jets, in the $KMR$, $LO \; MRW$ and
$NLO\;MRW$ schemes (i.e. the panels (a), (b) and (c), respectively),
demonstrating the contributions of the individual sub-processes. The
uncertainty bound is determined by manipulating the hard scale of
the $UPDF$, $\mu = E_{CM}/2$, by a factor of 2.  The panel (d)
represents a comparison between the results of the
$k_t$-factorization with the results from other frameworks, namely
the Balitsky-Kovchegov $TMD$ $PDF$ convoluted with running coupling
corrections ($rcBK$, see the references \cite{BK1,BK2}) and  the
Kutak-Sapeta  $TMD$ $PDF$ ($KS$), reference \cite{kutak5}.}
\label{fig9}
\end{figure}

After proving the success of our formalism in describing the
experimental data for the production of di-jets in the
forward-center rapidity region, we can move forward with the
prediction of a similar event, in the forward-forward sector, i.e.
by choosing the rapidity of the produced jets ($y_1$ and $y_2$) to
be both in the boundaries that where specified within the equation
(\ref{eq37}). Therefore, in the figure \ref{fig9} the reader is
presented with our predictions regarding the dependency of the
differential cross-section of the forward-forward di-jet production
($d\sigma_f/dp_t^f$) to the transverse momenta of the produced jets
($p_t$), in the framework of $k_t$-factorization. The panels (a),
(b) and (c) of the figure illustrate these predictions in the $KMR$,
the $LO\;MRW$ and  the $NLO\;MRW$ formalisms, respectively. The
contributions of the individual partonic sub-processes are included.
These contributions have the same general behavior as in the
forward-central case, in spite of the fact that the measured
contribution for the $g^{*}+g \to g+g$ and the $g^{*}+q \to g+q$
sub-processes are closer, compared to their counterparts from the
forward-center region,
\begin{eqnarray}
    &&\hat{\sigma}_{F-C}(g^{*}+g \to g+g)
    \gg \hat{\sigma}_{F-C}(g^{*}+q \to g+q)
    \gg \hat{\sigma}_{F-C}(g^{*}+g \to q+\bar{q}),
    \nonumber \\
    &&\hat{\sigma}_{F-F}(g^{*}+g \to g+g)
    \gtrsim \hat{\sigma}_{F-F}(g^{*}+q \to g+q)
    \gg \hat{\sigma}_{F-F}(g^{*}+g \to q+\bar{q}).
\end{eqnarray}
In addition, one can clearly perceive the effect of the $\Theta (
1-z-(k_t^2 / \mu^2) )$ constraint in the $NLO\;MRW$ results, causing
a steep descend in the corresponding histograms, in contrast with
the behaviors of the results of the $KMR$ and the $LO\;MRW$
formalisms. Again, the similarity of the predictions of the $KMR$
and the $LO\;MRW$ schemes are a consequence of our choice of the
hard scale, $\mu$. Such similarity was also observed else where,
e.g. the references \cite{Modarres7,Modarres8,Modarres9}, specially
in the smaller $x$ domains.

The panel (d) of the figure \ref{fig9} represents a comparison
between the results of the $k_t$-factorization with the results from
other frameworks, namely the $Balitsky$-$Kovchegov$ $TMD$ $PDF$
convoluted with the running coupling corrections ($rcBK$, see the
references \cite{BK1,BK2}) and  the $Kutak$-$Sapeta$  $TMD$ $PDF$
($KS$), the reference \cite{kutak5}. Both of these frameworks are
specially designed to describe the behavior of the small-x region,
incorporating the non-linear evolution of the $unintegrated$ parton
densities with the $KS$ framework and the high energy factorization
($HEF$) formalism, in accordance with the $BFKL$ iterative evolution
equation. In the absence of any experimental data, we refrain
ourselves from any assessments regarding these results.
Nevertheless, the predictions of the $KMR$ scheme (because of its
previous success) may provide a base line for a sound comparison.
Also, the singular behavior of the $NLO\;MRW$ results may appear
undesirable.

\begin{figure}[ht]
\includegraphics[scale=0.35]{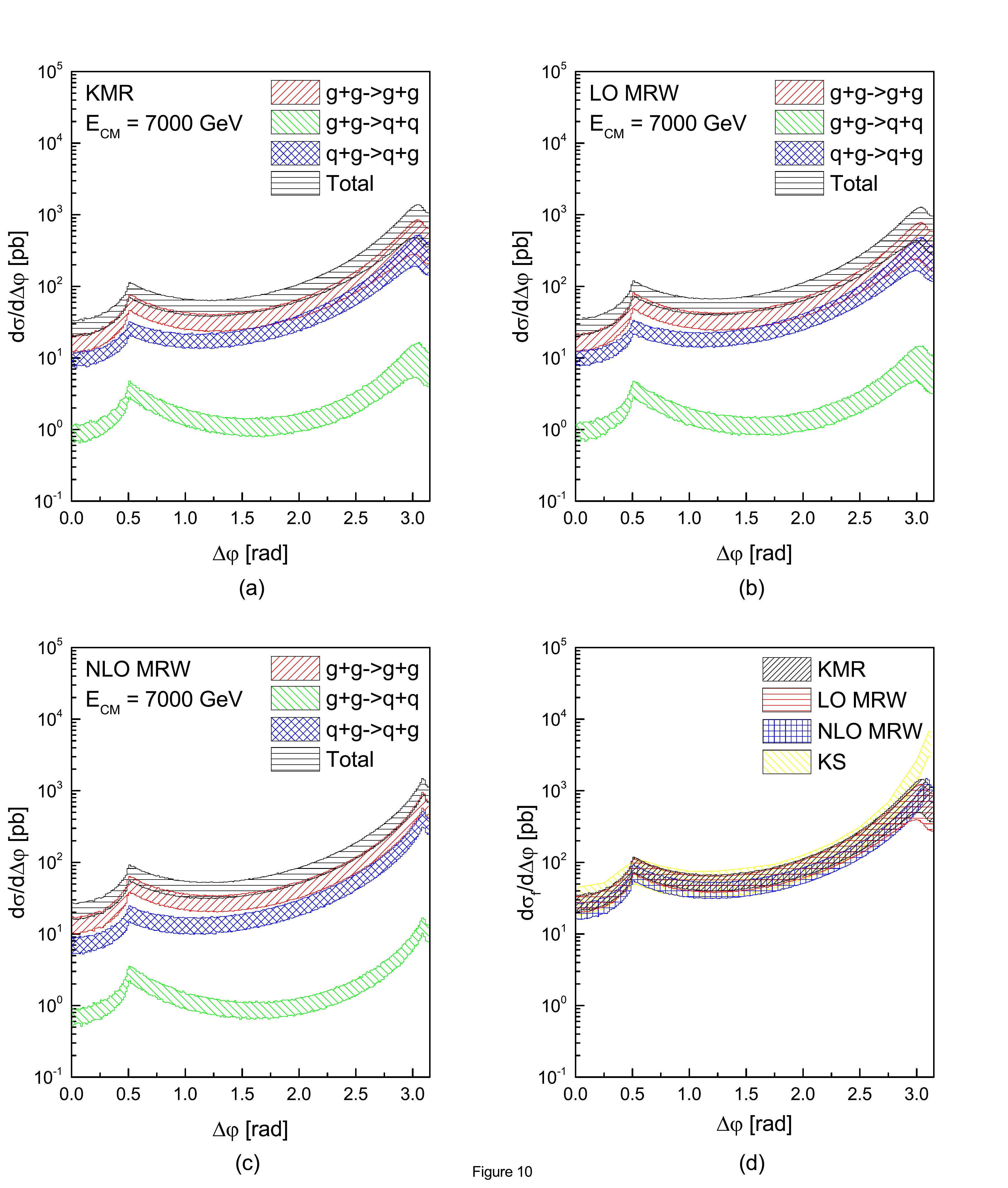}
\caption{The calculated predictions regarding the dependency of the
differential cross-section for the production of forward-forward
di-jets to $\Delta \varphi$ using the $UPDF$ of $k_t$-factorization
for $E_{CM}=7\;TeV$. The notion on the diagrams are as in the figure
\ref{fig9}. In the panel (d), we have compared our results with the
predictions made using the $KS$ $TMD$ $PDF$ from the reference
\cite{kutak5}.} \label{fig10}
\end{figure}

\begin{figure}[ht]
\includegraphics[scale=0.35]{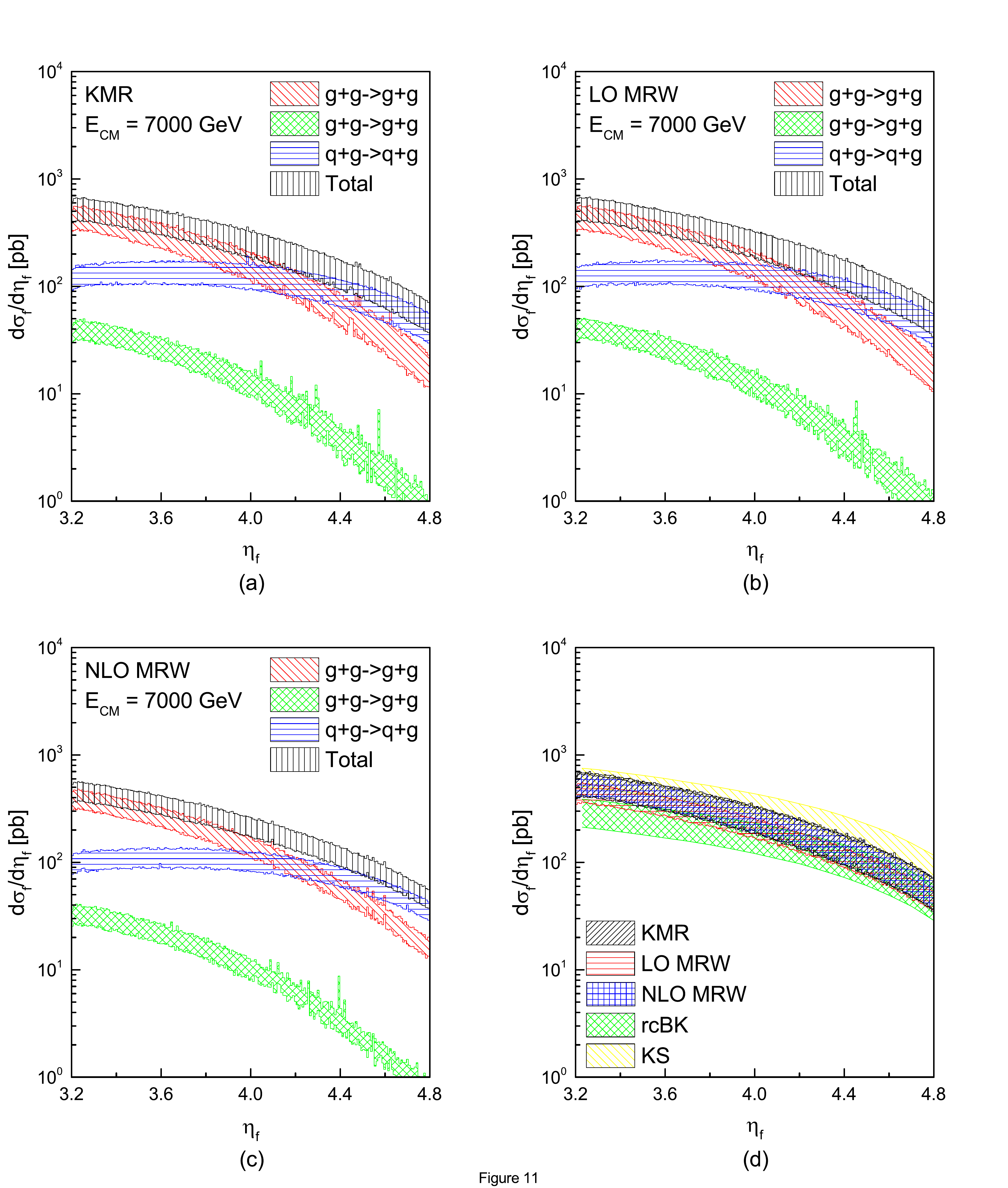}
\caption{The calculated predictions regarding the dependency of the
differential cross-section for the production of forward-forward
di-jets to rapidity of the produced jets, using the $UPDF$ of
$k_t$-factorization for $E_{CM}=7\;TeV$. The notion on the diagrams
are as in the figure \ref{fig9}. In the panel (d), we have compared
our results with the predictions made using the $rcBK$ and $KS$
$TMD$ $PDF$ from the reference \cite{kutak5}.} \label{fig11}
\end{figure}

Similar predictions are presented in the figures \ref{fig10} and
\ref{fig11}, describing the dependency of the differential
cross-section of the forward-forward di-jet production, to the angle
of the produced jets ($d\sigma_f/d \Delta \varphi$ to $\Delta
\varphi$ in the figure \ref{fig10}) and to their rapidity
($d\sigma_f/d \eta_f$ to $\eta_f$ in the figure \ref{fig11}). The
notions of these diagrams are as in the figure \ref{fig9}. The panel
(d) of each figure includes the comparison of the
$k_t$-factorization results to the existing results in the $rcBK$
and the $KS$ frameworks. The irregular behavior of the $NLO\;MRW$
scheme in both cases, manifests itself in the form of lower values
of the predicted differential cross-section. Again, the reliability
of these predictions lies within the excellent credit of the $KMR$
$UPDF$ in describing the high energy $QCD$ events.

In summary, throughout this work, we have tested the $UPDF$ of the
$k_t$-factorization, namely the $KMR$ and $MRW$ formalisms in the
$LO$ and the $NLO$, calculating the production rate of the di-jet
pairs at the deep inelastic $QCD$ collisions in the forward-center
rapidity sector, compared the results to the existing experimental
data of the $CMS$ collaborations and to the results of other
frameworks. {Through our analysis we have suggested that despite the
theoretical advantages of the $MRW$ formalism, the $KMR$ approach
performs as good as (if not better) behavior toward describing the
experimental data. This is in general agreement with our previous findings,
the references \cite{Modarres1, Modarres2, Modarres3, Modarres4,
Modarres5, Modarres6, Modarres7, Modarres8, Modarres9}.
Additionally, one can clearly see that the $KMR$ or $MRW$
prescription work better than the $KS$ in describing the experiment.
Based on these observations one concludes that the hard-scale
dependence should be necessarily included in $TMD$ analysis.
Furthermore, we have predicted the results of the similar events in
the forward-forward rapidity region, relying on the previous success
of the $UPDF$ of the $k_t$-factorization.}

\begin{acknowledgements}
$MM$ would like to acknowledge  the Research Council of University
of Tehran and Institute for Research and Planning in Higher
Education for the grants provided for him.

$MRM$ sincerely thanks N. Darvishi  for valuable discussions and
comments. $MRM$ extends his gratitude towards his kind hosts at the
Institute of Nuclear Physics, Polish Academy of Science for their
hospitality during his visit. He also acknowledges the Ministry of
Science, Research and Technology of Iran that funded his visit.
\end{acknowledgements}

\begin{appendix}
\section{The matrix elements of the partonic sub-processes}
Assuming that $\mu_1 = \mu_2 \equiv \mu$, the matrix element
squares, $|\mathcal{M}_{a_1+a_2 \to b_1+b_2}|^2$, corresponding to
the equations (\ref{eq19}) can be defined for a $QCD$ $IS$ event as
follows (also see the reference \cite{kutak5})
\begin{eqnarray}
    |\mathcal{M}_{g+g \to g+g}(x_1,\mu;x_2,k^2_{t})|^2 &&= C_1 A_1,
    \nonumber \\
    |\mathcal{M}_{g+g \to q+\bar{q}}(x_1,\mu;x_2,k^2_{t})|^2 &&= C_2 A_2 + C'_2 A'_2,
    \nonumber \\
    |\mathcal{M}_{q+g \to q+g}(x_1,\mu;x_2,k^2_{t})|^2 &&= C_3 A_3 + C'_3 A'_3,
    \label{A1}
\end{eqnarray}
with
\begin{eqnarray}
    C_1 &&= {9 \over 8} \left( {\alpha_S(\mu^2) \over 4\pi} \right)^2
    \nonumber \\[0.4cm]
    C_2 &&= {1 \over 6} \left( {\alpha_S(\mu^2) \over 4\pi} \right)^2 \; \; \; ,
    C'_2 = {1 \over 8} C_2,
    \nonumber \\[0.4cm]
    C_3 &&= {4 \over 9} \left( {\alpha_S(\mu^2) \over 4\pi} \right)^2 \; \; \; ,
    C'_3 = {1 \over 8} C_3,
\end{eqnarray}
and
\begin{eqnarray}
    A_1 &&= {2(e^{\Delta y} R_t+1)^2 (R_t e^{-\Delta y}(R_t e^{-\Delta y}+1)+1)^2 (cos(\Delta \varphi) + 2cosh(\Delta y))
    \over
    R_t^2(R_t e^{-\Delta y}+1)^2 (cos(\Delta \varphi) + cosh(\Delta y))},
    \nonumber \\[0.4cm]
    A_2 &&= { (R_t + e^{-\Delta y})^2 (R_t^2 e^{-\Delta y} + e^{\Delta y} )
    \over
    R_t  (R_t e^{-\Delta y}+1)^2},
    \nonumber \\[0.4cm]
    A'_2 &&={ (R_t + e^{-\Delta y})^2 (R_t^2 e^{-\Delta y} + e^{\Delta y})
    \over
    R_t  (R_t e^{-\Delta y}+1)^2 (cosh(\Delta y) - cos(\Delta \varphi))} cos(\Delta \varphi) ,
    \nonumber \\[0.4cm]
    A_3 &&=  { (R_t + e^{-\Delta y})^2 ((R_t + e^{\Delta y})^2 + R_t^2)
    \over
     2R_t (R_t e^{-\Delta y} + 1) (cosh (\Delta y) - cos(\Delta \varphi))} ,
    \nonumber \\[0.4cm]
    A'_3 &&= 2e^{-\Delta y} (e^{-\Delta y}-cos(\Delta \varphi)) { (R_t + e^{-\Delta y})^2 ((R_t + e^{\Delta y})^2 + R_t^2)
    \over
     2R_t (R_t e^{-\Delta y} + 1) (cosh (\Delta y) - cos(\Delta \varphi))},
\end{eqnarray}
where
$$ \Delta y = y_2 - y_1, \;\;\; R_t = {p_{1,t} \over p_{2,t}}.$$
Using the above information, one can calculate the cross-sections of
the equation (\ref{eq19}).
\end{appendix}

\end{document}